\documentclass[prl,showpacs,twocolumn,superscriptaddress]{revtex4}

\newcommand{\ket}[1]{|{#1}\rangle}

\newcommand{\D}[2]{\frac{\partial{#1}}{\partial{#2}}}

\usepackage{amsfonts}
\usepackage{amsmath}
\usepackage{graphicx}
\usepackage{amssymb}
\usepackage{color}
\usepackage{subfigure}
\usepackage[dvipsnames]{xcolor}

\usepackage{mathtools}

\begin{document}
\title{Curved spacetime from interacting gauge theories}

\author{Salvatore Butera}
\affiliation{INO-CNR BEC Center and Dipartimento di Fisica, Universit\`a di Trento, 38123 Povo, Italy}
\author{Niclas Westerberg}
\affiliation{SUPA, Institute of Photonics and Quantum Sciences, Heriot-Watt University, Edinburgh EH14 4AS, United Kingdom}
\affiliation{School of Physics \& Astronomy, University of Glasgow, Glasgow G12 8QQ, United Kingdom}
\author{Daniele Faccio}
\affiliation{School of Physics \& Astronomy, University of Glasgow, Glasgow G12 8QQ, United Kingdom}
\author{Patrik \"Ohberg}
\affiliation{SUPA, Institute of Photonics and Quantum Sciences, Heriot-Watt University, Edinburgh EH14 4AS, United Kingdom}

\begin{abstract}
Phonons in a Bose-Einstein condensate can be made to behave as if they propagate in curved spacetime by controlling the condensate flow speed. Seemingly disconnected to this, artificial gauge potentials can be induced in charge neutral atomic condensates by for instance coupling two atomic levels to a laser field. Here we connect these two worlds and show that synthetic interacting gauge fields, i.e., density-dependent gauge potentials,  induce a non-trivial spacetime structure for the phonons. Despite the creation of effective horizons for phonons solely depends on the flow speed of the condensate, this allows for the creation of new spacetime geometries which can be easily designed by tuning the transverse laser phase. By exploiting this new degree of freedom we show that effectively charged phonons in 2+1 dimensions can be simulated, which behave as if they move under the influence of both a gravitational and an electromagnetic field.
\end{abstract}


\maketitle

{\it Introduction.} Analogue gravity has in the recent decades provided us with a powerful simulator of quantum fields in curved spacetime. In this discipline, we seek physical systems whose underlying dynamics might be different from gravity, but where an apparent spacetime picture emerges for some field in the system \cite{barcelo2005analogue,novello2002artificial,faccio2013analogue}. Consequently, a quantum field would then behave as though it is in curved spacetime. In this way, it is possible to study physics that relies on the spacetime structure, but not the gravitational dynamics. Thus, by analogy, effects such as Hawking radiation, the process in which black holes evaporate by emitting thermal radiation, proposed by S Hawking \cite{hawking1974black,hawking1975particle}, the related Unruh effect \cite{unruh1981experimental}, and cosmological particle creation \cite{schro1939proper,parker1968particle} are no longer unattainable experimentally. For instance, this has enabled the experimental study of the Hawking effect in flowing Bose-Einstein Condensates (BEC) \cite{steinhauer2015observation,steinhauer2014observation,lahav2010realization}, flowing water \cite{rousseaux2008observation} and nonlinear optics \cite{rubino2011experimental}, although in the latter the observed radiation might have had a more complicated origin. In-depth theoretical studies in BECs \cite{visser1998acoustic,reznik2000origin,garay2000sonic,leonhardt2003bogoliubov,leonhardt2003theory,liberati2006analogue,barcelo2006quasi,finazzi2010black}, ultracold fermions \cite{giovanazzi2005hawking}, moving/nonlinear optical media \cite{leonhardt1999optics,leonhardt2000relativistic,unruh2003slow,petev2013blackbody}, rings of trapped ions \cite{horstmann2010hawking} and superconducting circuits \cite{nation2009analogue} to name a few, has exposed a multitude of properties of analog Hawking radiation. For example, this unveiled the existence of density-density correlations in a BEC with a sub/supersonic transition \cite{balbinot2008nonlocal} that are vital for experimental investigation \cite{steinhauer2015observation}. Another example is the deep connection between seemingly unrelated quantum field phenomena such as the dynamical Casimir effect,  Hawking radiation and time-refraction  \cite{mendonca2008vacuum,nation2012colloquium,schutzhold2011refractive}.  Although Hawking radiation has held most of the attention of the community, analogue gravity is not restricted to this phenomenon alone. For example cosmological particle creation was studied in \cite{gibbons2003fedichev,analog2007jain,faccio2013analogue,westerberg2014experimental,Prain2017}.

However, so far analogue gravity has been restricted to the study of spacetimes created using only \textit{one} degree of freedom \cite{barcelo2005analogue,novello2002artificial,faccio2013analogue}. For instance, the current of a BEC has as of yet been the sole determiner of the analogue spacetime felt by the phonons propagating in it. In this article, we will introduce a \textit{second} degree of freedom in the form of a gauge potential, by which richer physics can in principle be simulated. These synthetic gauge potentials emerge in cold-atom systems by for instance coupling  the centre-of-mass motion of the particles to their internal degrees of freedom with a laser \cite{dalibard_2011,goldman_2013}. A Berry connection \cite{berry_1984} then emerges in the form of a vector potential for the centre-of-mass dynamics. The coupled atomic states must be long-lived compared to the characteristic time scales of the investigated analogue phenomena. Atomic losses due to spontaneous decay, happening on competing time scales, would overwhelm the sought physics and eventually destroy the condensate. Working with atomic optical transitions, promising candidates could for instance be Ytterbium and Strontium which have extremely long lived states of the order of seconds \cite{dalibard_2011}. An alternative route in order to avoid spontaneous emission and heating is to use dark states and three-level atoms \cite{theuer_1999,dalibard_2011}.

In this paper we start from a density-dependent gauge potential, first described in \cite{edmonds_2013a}, and demonstrate how this adds a new degree of freedom to analogue gravity models. We show that the nonlinearity of the vector potential introduces a new mechanism for designing effective spacetime geometries in a condensate. We then discuss how the additional degree of freedom introduced by the nonlinear gauge potential allows us to extend analogue gravity beyond what is currently possible. In particular we show that effectively charged particles in $2+1$ dimensions can be simulated with phonons, which behaves as particles subjected to both a gravitational and an electromagnetic field.

\emph{The physical system.} We consider for simplicity a BEC composed of two-level atoms collisionally interacting with each other. The internal bare states $\ket{1}$ and $\ket{2}$ are dressed because of the interaction with an external laser beam, and we call the resulting eigenstates $\ket{\pm}$. In previous works \cite{edmonds_2013a,edmonds_2015,zheng_2015,butera2015quantized}, a density-dependent effective vector potential $\mathbf{A}$, and a scalar potential $W$, have been shown to emerge acting on the atoms as a result of collision-induced shifts of the energy of the dressed states. An explicit expression for these potentials can be obtained by working in the weakly interacting limit, where the strength of the atom-atom interaction  is much smaller than the characteristic energy of the laser-matter coupling. In such a regime the interatomic interaction can be treated as a small perturbation to the light-atom interaction. We consider here the simplest case of a light field perfectly resonant with the atomic transition. Up to first order in the perturbation theory, the effective vector and scalar potentials take the form \cite{edmonds_2013a,butera2015quantized}
\begin{align}
	\mathbf{A}_\pm & =\mathbf{A}^{(0)}\pm \mathbf{a}_1 \rho_\pm(\mathbf{r}), \label{Vec_Pot}\\
	W & =\frac{\left|\mathbf{A}^{(0)}\right|^2}{2m}, \label{Scal_Pot}
\end{align}
where the subscript $\pm$ refers to atoms in one or the other dressed state. Here we indicated by $\mathbf{A}^{(0)}=-\frac{\hbar}{2}\boldsymbol{\nabla}\phi$ the single particle component of the vector potential, $\phi$ the phase of the laser beam, and $\rho_\pm(\mathbf{r})=|\psi_\pm|^2$ the density of  the dressed state, with $\psi_\pm$ the condensate order parameter. The vector $\mathbf{a}_1=\left[(U_{11}-U_{22})/(8\Omega)\right]\,\boldsymbol{\nabla}\phi$ controls the direction and strength of the first order nonlinear, density-dependent contribution, in which $\Omega$ is the Rabi frequency, that sets the strength of the interaction. The mean-field coupling constants are defined by $U_{ij}=4\pi\hbar^2 a_{ij}/m$, where the scattering lengths $a_{ij}$ $(i,j=1,2)$ describe the interaction strength between atoms in the different collisional channels. The zero-order term $\mathbf{A}^{(0)}$ can be removed by the gauge transformation $\psi_\pm\rightarrow e^{i\phi(\mathbf{r})/2}\psi_\pm$ provided the corresponding magnetic field is zero everywhere. In this situation, only the nonlinear term of the vector potential is non trivial, and the Gross-Pitaevskii equation (GPE) for the mean-field components of the system take the form\cite{edmonds_2013a,butera2015quantized},
\begin{equation}
	i\hbar\frac{\partial\psi_\pm}{\partial t}=\left[\frac{\left(\mathbf{p}-\mathbf{A}_\pm\right)^2}{2m}\mp \mathbf{a}_1\cdot \mathbf{j}+W+U\rho_\pm\right]\psi_\pm
\label{GP}
\end{equation}
where $U=(U_{11}+U_{22}-2U_{12})/4$ and the current is defined as
\begin{equation}
	\mathbf{j}=\frac{\hbar}{2mi}\left[\psi_\pm^*\left(\boldsymbol{\nabla}-\frac{i}{\hbar} \mathbf{A}_\pm\right)\psi_\pm-\psi_\pm\left(\boldsymbol{\nabla}+\frac{i}{\hbar} \mathbf{A}_\pm\right)\psi_\pm^*\right].
\label{Current}
\end{equation}
For definiteness, and without loss of generality for the following arguments, we work hereafter with the $(+)$ component of the condensate, and drop the subscript in the quantities defined above. Had we instead chosen to work with the $(-)$ component of the condensate, the following arguments would have been the same provided we exchange $U_{11}\leftrightarrow U_{22}$.

\emph{The effective metric.} The GPE in Eq.~\eqref{GP} provides the framework from which the effective spacetime felt by phonons can be derived. To this aim, we work in the hydrodynamic formalism, and write the order parameter in terms of the particle density $\rho$ and its phase $S$, as $\psi=\sqrt{\rho}e^{i S}$. In terms of these quantities, the Eq.~\eqref{GP} is equivalent to the continuity and the (quantum) Euler equations
\begin{equation}
 \D{\rho}{t}+\boldsymbol{\nabla}\cdot\left(\rho\mathbf{v}\right)=0\label{syst_cont}
 \end{equation}
and
\begin{align}
\hbar\D{S}{t} &=\frac{\hbar^2}{2m}\frac{\boldsymbol{\nabla}^2\sqrt{\rho}}{\sqrt{\rho}}-\frac{\hbar^2}{2m}(\boldsymbol{\nabla} S)^2 \nonumber \\ 
&\quad\quad + \quad 2\rho\mathbf{a}_1\cdot \mathbf{v}+\rho\left(\frac{\rho\left|\mathbf{a}_1\right|^2}{2m}-U\right).\label{syst_Euler}
\end{align}
Here $\mathbf{v}= ({\hbar}/{m}) \boldsymbol{\nabla} S-{\mathbf{A}}/{m}$ is the physical velocity in the condensate. Phonons represent long wavelength excitations of the system above its mean-field component. At the classical level, the dynamics of these excitations can be deduced by linearizing Eqs.~\eqref{syst_cont} and \eqref{syst_Euler} above the condensate background. By indicating with $\rho_0$ and $S_0$ respectively the density of particles in the condensate and the phase of the order parameter, we include small perturbations in the theory by writing the density and the phase as $\rho=\rho_0+\rho_1$ and $S=S_0+S_1$, respectively. Here $\rho_1$ and $S_1$ account for small deviations from the condensate component. Retaining only the first order terms in $S_1$, and $\rho_1$, {and working in the hydrodynamic regime in which the quantum pressure can be neglected, the dynamics of the excitations is described by the equation}
\begin{equation}
	\Delta S_1\equiv \frac{1}{\sqrt{-g}}\frac{\partial}{\partial x^\mu}\left(\sqrt{-g}g^{\mu\nu}\frac{\partial S_1}{\partial x^\mu}\right)=0.
\label{dAlembert}
\end{equation}
This is the equation for a scalar field propagating in a curved spacetime \cite{birrell1984quantum} and here describes the evolution of acoustic fluctuations living in the effective background geometry described by the metric $g_{\mu\nu}$ tensor
\begin{equation}
	g_{\mu\nu}=\frac{\rho_0}{c_s}\left(\begin{array}{cc}
        -\left(c^2-v_0^2\right) & -\left(v_0\mp v_a\right)^j \\ 
        -\left(v_0\mp v_a\right)^i & \delta^{ij}
\end{array}\right).
\label{Metric_offdiag}
\end{equation}
Here $\mathbf{v}_0=\hbar\boldsymbol{\nabla} S_0/m-\mathbf{v}_a$ is the zeroth order component of the velocity field in the condensate, $\mathbf{v}_a=\rho_0\mathbf{a}_1/m$ is the effective velocity induced by the nonlinear vector potential, $c_s^2=g' n_0/m=c^2- 2 \mathbf{v}_a\cdot \mathbf{v}_0 +v_a^2$ is the local speed of sound in the condensate (with $v_a=\left|\mathbf{v}_a\right|$), while $c^2=g n_0/m$ is the value it would take in absence of the potential. Eq.~\eqref{Metric_offdiag} shows that cross terms in the metric, mixing the space and time coordinates in the laboratory reference frame, are induced by the nonlinear vector potential. Interestingly, the time-time component (as well as the pure spatial components) of the metric is not affected by the nonlinear potential, and solely the physical velocity of the condensate is responsible for the establishment of an ergo-region and eventually the appearence of acoustic horizons in the system. The nonlinear potential thus provides an extra degree of freedom that can be exploited in order to design effective spacetimes for phonons. In particular, Eq.~\eqref{Metric_offdiag} reveals that a nontrivial curved spacetime can be induced even for a static condensate, for which the physical velocity $\mathbf{v}_0$ of particles is zero. Moreover, effects such as cosmological particle creation, or dynamical Casimir effect, which is the particle creation process triggered by the parametric amplification of the vacuum fluctuations in the presence of time-dependent boundary conditions, can be relatively easily implemented by simply modulating in time the light-matter interaction parameters, such as the Rabi frequency or the detuning.

We show now that the new degree of freedom provided by the nonlinear vector potential can be used to simulate the dynamics of charged particles in two-dimensions, moving under the combined influence of both a gravitational and an electromagnetic field.

{\emph{Simulating effectively charged phonons.}} We start with the metric in Eq.~\eqref{Metric_offdiag} and make the following time-coordinate transformation
\begin{equation}
	dt'=dt+\frac{\left(\mathbf{v}_0-\mathbf{v}_a\right)}{c^2-v_0^2}\cdot d\mathbf{x}.
\label{CoordTransform}
\end{equation}
This is a type of co-moving coordinates, and in this new set of coordinates, the spacetime curvature has been pushed from the time-space into the space-space components of the metric. Disregarding the conformal factor, the effective metric seen by phonons now reads
\begin{multline}
	ds^2=-\left(c^2-v_0^2\right)dt'^2\\
	+\left[\delta_{ij}+\frac{\left(v_0-v_a\right)_i}{\sqrt{c^2-v_0^2}}\frac{\left(v_0-v_a\right)_j}{\sqrt{c^2-v_0^2}}\right]dx_i dx_j.
\label{MetricTransformed}
\end{multline}
We work in the limit $\varphi^2\equiv{\left(v_0-v_a\right)_3^2}/(c^2-v_0^2)\gg 1$ and call $\mathcal{A}_i\equiv\left(v_0-v_a\right)_i/\left(v_0-v_a\right)_3$ $(i=1,2)$. Hereafter we relabel the full $3+1$ metric as $g_{MN}$, with the indexes indicated by capital Roman letters spanning the full four-dimensional manifold $M,N,...=0,1,2,3$, and define the reduced $2+1$-dimensional metric $h_{\mu\nu}=\text{diag}\left[-\left(c^2-v_0^2\right),1,1\right]$ (with the Greek indexes taking the values $\mu,\nu,...=0,1,2$). In terms of these quantities, the full $3+1$ metric can be written as
\begin{equation}
	g_{MN}=\left(\begin{array}{cc}
         h_{\mu\nu}+\varphi^2 \mathcal{A}_\mu \mathcal{A}_\nu & \varphi^2 \mathcal{A}_\mu \\ 
          \varphi^2 \mathcal{A}_\mu & \varphi^2
\end{array}\right).
\label{MetricA2}
\end{equation}
Such a metric has the same form as the ansatz introduced by Klein \cite{Klein} in the context of dimensional reduction in the Kaluza-Klein (KK) theory. Such a theory represents the first attempt of pursuing the unification of electromagnetism and gravity, in the framework of a more general theory of gravity in higher dimensions \cite{Bailin1987,Kaluza,Klein}. The simpler implementation of this idea postulates the existence of a hidden fifth dimension, whose characteristic scale is of the order of the Planck length, and generalizes the Einstein theory of gravity to this higher dimensional spacetime. Electromagnetism then emerges upon dimensional reduction of the theory, i.e. ``compactifying'', or in other words integrating out, the extra dimension on a circle \cite{Bailin1987}.
 
By assuming that the dilaton field $\varphi$ is constant, as well as the $x_3$-independence of the metric $h_{\mu\nu}$ and the vector potential $\mathcal{A}_\mu$, the effective metric Eq.~\eqref{MetricA2} can be seen as a KK metric in a $2+1$ physical spacetime, in which the role of the extra dimension is played by the third space component, labelled by the coordinate $x_3$. Following this reasoning, $\mathcal{A}_\mu$ takes the role of an effective electromagnetic potential for phonons in the $2+1$ spacetime. Phonons in the $x_1-x_2$ plane of the condensate thus behave as charged particles subject to both a gravitational and an electromagnetic field.

Strictly speaking, the analogue model we developed is equivalent to the KK theory up to the conformal factor in the metric in Eq.~\eqref{Metric_offdiag}. The full equivalence is attained in the case such a factor is constant in time and homogeneous in space. Given the expression above for the sound velocity $c_s$, this is fulfilled if the following conditions are satisfied: i) the density $\rho_0$ in the system is constant, ii) the physical velocity $\mathbf{v}_0$ is orthogonal to the induced velocity $\mathbf{v}_a$, and iii) $v_a^2\ll c^2$. In this case $c_s$ reduces to $c$, that is constant in virtue of (i). In order to demonstrate the emergence of an effective charge for phonons, we start from the action
\begin{equation}
S=-\frac{1}{2}\int{d^4x\sqrt{-G}g^{MN}\left(\partial_M S_1\right)\left(\partial_N S_1\right)},
\label{Action1}
\end{equation}
from which the wave equation in Eq.~\eqref{dAlembert} can be deduced. The factor $G$ here is the determinant of the metric tensor in Eq.~\eqref{MetricA2}, which can also be expressed in terms of the determinant of the $2+1$ metric tensor $h=\text{det}\left[h_{\mu\nu}\right]$ as
\begin{equation}
\begin{split}
G&=\text{det}\left[G_{MN}\right]\\
&=\text{det}\left[h^{\mu\nu}\right]^{-1}\left[\left(\frac{1}{\varphi^2}+\mathcal{A}^2\right)-\mathcal{A}^\mu h_{\mu\nu} \mathcal{A}^\nu\right]^{-1}\\
&=h\varphi^2
\end{split}
\label{Action2}
\end{equation}
Since the metric in Eq.~\eqref{MetricA2} does not depend on the coordinate $x_3$ by assumption, we can expand $S_1$ in terms of the mode basis $\{\xi_n(x_3)\}$ in the $x_3$ direction
\begin{equation}
S_1=\sum_n{s_n\left(x^\nu\right)\xi_n(x_3)}.
\end{equation}
We consider for simplicity periodic boundary conditions and write the eigenmodes as $\xi_n(x_3)=e^{ik_3^n x_3}/\sqrt{L}$, where $L$ is the length of the system in the $x_3$ direction. By inserting this expansion into Eqs.~\eqref{Action1} and integrating over $x_3$, we find the reduced action in the form
\begin{align}
\mathcal{S} &=\varphi\sum_n\bigg\{-\frac{1}{2}\int d^3x  \\
&\sqrt{-h}\bigg[\left(h^{\mu\nu}\left[\left(\partial_n-i k_3^n \mathcal{A}_\mu\right)s_n\right]^2\right)-\left(\frac{k_3^n}{\varphi}\right)^2s_n^2\bigg]\bigg\}. \nonumber
\label{Action4}
\end{align}
It is given by the sum of an infinite number of actions, describing the dynamics of massive charged particles in the 2+1 dimensional transverse $x_1-x_2$ plane, and is what is called in literature the \emph{Kaluza-Klein tower}. The values of the effective charges and masses for each mode are set by the value of the momentum of the mode in the $x_3$ direction, that is $q_n=\hbar k_3^n$ $m_n=\hbar k_3^n/c\varphi$. The ratio $q/m=c\varphi$, which is related to the cyclotron frequency $\omega_c = Bq/m$, is however the same for all the modes.

{\emph{Experimental considerations.}} For an experimental realisation of the analog gravity effects discussed above, a number of criteria must be fulfilled. Firstly, the density-dependent gauge potential relies on the Rabi frequency and the corresponding energy scale to dominate over any collisional interaction energies. In practice this means $\hbar\Omega$ must be larger than the chemical potential $\mu$ of the Bose-Einstein condensate. Secondly, we need to ensure a suitable choice of atomic states and scattering lengths, such that a non-zero current nonlinearity can be obtained. For this one needs $U_{11}\ne U_{22}$  which, if not readily available, can be achieved using Feshbach resonances. In other words, one needs to be in the adiabatic regime, where the dressed states arising from the light-matter interaction are not coupled. This not only requires that $\hbar\Omega\gg\mu$, but the atomic states must also be long lived. Finally, the nonlinear gauge potential should take some specific spatial form in order to emulate a specific spacetime structure in Eq.~\eqref{Metric_offdiag} or effective vector potential in Eq.\eqref{MetricTransformed}. Similarly to the proposal in Ref.~\cite{marino2008acoustic}, the relevant phase profiles required to this purpose can be easily obtained using standard beam shaping technologies. 
\begin{figure}[t]
\centering
\includegraphics[width=0.8\linewidth,angle=0]{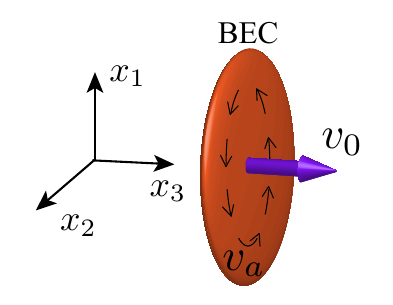}
\caption{Scheme of the experimental setup discussed in the text, needed to simulate effectively charged phonons in a BEC. A pancake-shaped condensate is let to move along the $x_3$ direction with a constant velocity $\mathbf{v}_0$. The internal states of the two-level atoms are coupled by a laser beam, whose phase gradient $\phi$ is circularly directed.}
\label{fig:setup}
\end{figure}

Regarding the implementation of effectively charged phonons, a possible setup useful for the purpose is illustrated in Fig.~\ref{fig:setup}. Here we have a 2D condensate moving in the $x_3$ direction with a constant velocity $v_0$. We assume the phase gradient of the laser lying in the transverse $x_1-x_2$ plane, where the effective dynamics for phonons is expected to appear. In these conditions we thus obtain $\varphi^2=v_0^2/\left(c^2-v_0^2\right)$ and $\mathcal{A}_i=-v_{a,i}/v_0$ (with $i=1,2$). In order to fulfil the condition $\varphi^2\gg 1$, we assume the velocity of the condensate approaching the speed of sound $v_0\approx c$. We take the gradient of the laser phase $\boldsymbol{\nabla}\phi$ directed along the angular direction $\hat{e}_\theta$. In order to clearly separate the mean-field dynamics from the dynamics of phonons, we tailor the laser beam in such a way that $\boldsymbol{\nabla}\phi\sim (1/r) \hat{e}_\theta$. In this case, the zero-order synthetic vector potential gives rise to a zero effective magnetic fields for the atoms. We choose the remaining experimental parameters, that is the Rabi frequency, and the mean-field coupling constants $U_{11}-U_{22}$, in such a way that $\boldsymbol{\mathcal{A}}=(B/2) r\,\hat{e}_\theta$ so that the corresponding effective magnetic field felt by phonons is constant and equal to $B$ (note here that, if we let $U_{11}-U_{22}$ vary in space, we need to suitably shape $U_{12}$ as well, in order to ensure that the speed of sound $c_s$ is homogeneous and the conformal factor in the metric \eqref{Metric_offdiag} constant). With this choice of the set-up, the coordinate transformation in Eq.~\eqref{CoordTransform} can be integrated giving
\begin{equation}
	t'=t+\frac{v_0}{\left(c^2-v_0^2\right)}\left(z+\frac{B}{2}r^2\theta\right).
\label{CoordTransformIntegr}
\end{equation}
In order for the phonons to gain an effective mass and charge, we perturb the condensate in the longitudinal direction $x_3$. We then imprint an asymmetric perturbation in the transverse $x_1-x_2$ plane and look for the precession of the small amplitude excitations with the cyclotron frequency $\omega_c$. We finally note that, with the setup here proposed, the reduced $2+1$ metric $h_{\mu\nu}$ is flat. The effective dynamics of phonons is thus that of a charged particle in a Minkowskian spacetime. More interesting physics can be simulated by exploiting the other degree of freedom at disposal, that is the velocity $\mathbf{v}_0$ of the condensate, which we assumed constant in the previous discussion. By making such a velocity space (in the transverse plane) or time dependent, the effective metric for the phonons is effectively curved. In this cases it might be experimentally difficult to ensure that the conformal factor is homogeneous throughout space. In the geometrical acoustic limit (or in other terms in the eikonal approximation) \cite{barcelo2005analogue} however, which is insensitive to the conformal factor, the physics discussed here can still be explored.

\emph{Summary and Conclusion.} We have shown that density-dependent synthetic gauge fields acting on neutral atoms can be exploited in order to enrich the physics that can be simulated in a BEC implementation of gravity analogues. Such nonlinear fields open up the possibility of introducing a new, independent, and versatile degree of freedom, useful for designing effective spacetime for phonons. The structure of this effective spacetime depends on the details of the atom-field interaction parameters and can be easily adjusted by opportunely designing the experimental setup. As an application of the model, we showed that effectively charged particles subjected to both a gravitational and an electromagnetic field can be simulated with phonons. We discussed the experimental implementation of the model, pointing out that such analogue physics can be attained even with a static condensate, using exclusively the degree of freedom provided by the nonlinear synthetic vector potentials acting on the atoms.

\acknowledgements
S.B gratefully acknowledge financial support by the Julian Schwinger Foundation. N.W acknowledges support from the EPSRC CM-CDT Grant EP/L015110/1, P.\"O acknowledges support from EPSRC EP/M024636/1. D.F. acknowledges financial support from the European Research Council under the European Unions Seventh Framework Programme (FP/20072013)/ERC GA 306559 and the EPSRC (Grant EP/P006078/2). 


\end{document}